%% file: main.tex
\documentclass[%
 aip,jcp,superscriptaddress,
 amsmath,amssymb,floatfix,longbibliography,
 reprint,%
]{revtex4-2}
\usepackage{nicematrix}
\usepackage[utf8]{inputenc}
 \usepackage[T1]{fontenc}
\usepackage{chngcntr}
\usepackage{graphicx}
\usepackage{dcolumn}
\usepackage{bm}
\usepackage{xcolor}
\usepackage[normalem]{ulem}
\usepackage{physics}
\usepackage{chemformula}
\usepackage{soul}
\usepackage{subcaption}
\usepackage{booktabs}

\let\ket=\pket 

\usepackage{braket}
\usepackage{tikz}
\usetikzlibrary{external}
\usepackage{float}
\usepackage[colorlinks]{hyperref}
\hypersetup{ 
    colorlinks=true,
    linkcolor=red!80!black,
    urlcolor=magenta!80!black,
    citecolor=green!70!black,
}
\usepackage{cleveref}
\usepackage{ragged2e}
\usepackage{nameref} 

\newcommand{\str}[1]{}
\newcommand{\newb}[1]{#1}

\DeclareMathOperator*{\argmin}{arg\,min}
\begin{document}
\preprint{}

\title{Multidimensional quantum dynamics with Explicitly Correlated Gaussian wave packets using Rothe's method}

\author{Simon Elias Schrader}
\email{s.e.schrader@kjemi.uio.no}
\author{Thomas Bondo Pedersen}%
\author{Simen Kvaal}%

\affiliation{Hylleraas Centre for Quantum Molecular Sciences, Department of Chemistry, University of Oslo, P.O. Box 1033 Blindern, N-0315 Oslo, Norway 
}%

\date{\today}

\begin{abstract}\small
    In a previous publication [J. Chem. Phys., \textbf{161}, 044105 (2024)], it has been shown that Rothe's method can be used to solve the time-dependent Schrödinger equation (TDSE) for the hydrogen atom in a strong laser field using time-dependent Gaussian wave packets. Here, we generalize these results, showing that Rothe's method can propagate arbitrary numbers of thawed, complex-valued, Explicitly Correlated Gaussian functions (ECGs) with dense correlation matrices for systems with varying dimensionality. We consider the multidimensional Henon-Heiles potential, and show that the dynamics can be quantitatively reproduced using only 30 Gaussians in 2D, and that accurate spectra can be obtained using 20 Gaussians in 2D and 30 to 40 Gaussians in 3D and 4D. Thus, the relevant multidimensional dynamics can be described at high quality using only a small number of ECGs that give a very compact representation of the wave function. This efficient representation, along with the demonstrated ability of Rothe's method to propagate Gaussian wave packets in strong fields and ECGs in complex potentials, paves the way for accurate molecular dynamics calculations beyond the Born-Oppenheimer approximation in strong fields.
\end{abstract}

\maketitle

\section{Introduction}\label{sec:Introduction}
\input{intro}

\section{Theory}\label{sec:Methods}

\subsection{Explicitly Correlated Gaussian wave packets}
\newb{Following Ref. \citenum{Mitroy_Gaussian2013}, we refer to ECGs as Gaussian functions that have explicit pair correlation terms, $r_i\cdot r_j$, in the exponent. }A $D$-dimensional ECG is a wave function of the form
\begin{align}\label{eq:Gaussian}
    g_m&=g(d_m,\boldsymbol{\alpha}_m)\nonumber \\
    &=d_m\exp\left(-\left(\boldsymbol{r}-\boldsymbol{\mu}_m\right)^T\boldsymbol{A}_m\left(\boldsymbol{r}-\boldsymbol{\mu}_m\right)\right),
\end{align}
where the weight $d_m$ is a complex number containing information about phase and norm, $\boldsymbol{A}_m$ is a complex-symmetric $D\times D$ matrix with positive definite real part, $\Re(\boldsymbol{A}_m)>0$ guaranteeing square-integrability, and $\boldsymbol \mu_m$ is a complex vector of length $D$. One can generally parameterize $\boldsymbol{A}_m$ as 
\begin{equation}\label{eq:ECG}
    \boldsymbol{A}_m=\boldsymbol{L}_m\boldsymbol{L}_m^T + \text{i}(\boldsymbol{K}_m+\boldsymbol{K}_m^T),
\end{equation}
where both $\boldsymbol{L}_m$ and $\boldsymbol{K}_m$ are real, lower triangular matrices. The parameterization of the real part of $\boldsymbol{A}_m$ as a Cholesky decomposition ensures symmetry and positivity, while the imaginary part is symmetric. Thus, every Gaussian is represented by two real lower triangular matrices and two real vectors $\Re(\boldsymbol{\mu}_m)$ and $\Im(\boldsymbol{\mu}_m)$, resulting in $D(D+3)$ real, nonlinear parameters per Gaussian ($10$ in $2D$, $18$ in $3D$, $28$ in $4D$). In the following, we generally collect the nonlinear parameters in the vector $\boldsymbol{\alpha}_m$.

An equivalent parameterization of an ECG is the Heller form, \cite{Heller_1976,lubich2008quantum} which reads
\begin{align}\label{eq:ECG_Heller}
    \tilde g_m&=\tilde g(\boldsymbol C_m,\boldsymbol p_m,\boldsymbol q_m,\zeta_m) \nonumber \\
&=\exp \Bigg(\text{i}\Big(\frac{1}{2}(\boldsymbol{r}-\boldsymbol{q}_m)^T \boldsymbol{C}_m(\boldsymbol{r}-\boldsymbol{q}_m) \nonumber \\
    &\qquad\qquad    + \boldsymbol{p}_m^T (\boldsymbol{r}-\boldsymbol{q}_m)+\zeta_m\Big)\Bigg),
\end{align}
where $\boldsymbol{q}_m, \boldsymbol{p}_m$ are real vectors of length $D$, $\zeta_m$ is a complex number containing information about phase and norm, and $\boldsymbol{C}_m$ is a complex symmetric matrix with positive definite imaginary part. The two forms (eq. \eqref{eq:ECG} and eq. \eqref{eq:ECG_Heller})  can easily be transformed from one into the other. 

In this paper, we use ECGs in the form of \cref{eq:ECG} as basis functions for our time-dependent calculations. For numerical reasons, the absolute value of $d_m$ is chosen such that each ECG is normalized.
\subsection{Rothe's method for the time-dependent Schr{\"o}dinger equation}\label{sec:Rothe}
We present here a short summary of Rothe's method, referring the reader to Ref. \citenum{schrader2024time} for an in-depth presentation.\\
The time dependent Schrödinger equation in units with $\hbar=1$, and omitting the spatial dependence of the wave function for notational convenience, reads
\begin{equation}
    \text{i}\frac{\partial}{\partial t}\Psi(t)=\hat{H}(t)\Psi(t).
\end{equation}
Applying the Crank-Nicolson propagator with step size $\Delta t$ to the wave function  at time $t_i$, the wave function at the time $t_{i+1}=t_i+\Delta t$ reads \cite{joachain}
\begin{equation}
\label{eq:crank-nicolson}
    \Psi(t_{i+1})=\hat A_{i}^{-1} \hat A_{i}^\dagger\Psi\left(t_i\right),
\end{equation}
where
\begin{equation}
    \hat A_{i}=\hat I+\text{i}\frac{\Delta t}{2}\hat H\left(t_i+\frac{\Delta t}{2}\right).
\end{equation}
This can be variationally reformulated as 
\begin{equation}\label{eq:Rothe}
    \Psi(t_{i+1})=\argmin_{\chi}\norm{\hat A_i\chi-\hat A_i^\dagger\Psi(t_i)}^2.
\end{equation}
This is Rothe's method. \cite{deuflhard2012adaptive,kvaal2023need} Parameterization of the wave function in a non-linear basis consisting of $N(t)$ basis functions with linear coefficients $\boldsymbol{c}$ and nonlinear coefficients $\boldsymbol{\alpha}$ 
\begin{equation}\label{eq:WF_ansatz}
    \chi(t)=\sum_{m=1}^{N(t)} c_m(t)\phi_m(\boldsymbol{\alpha}(t)),
\end{equation}
and insertion in the ansatz yields an optimization problem for the optimal parameters describing $\Psi(t_{i+1})$:
\begin{equation}\label{eq:Rothe_opt}
\boldsymbol{\alpha}_{\text{opt.}}^{i+1},\boldsymbol{c}_{\text{opt.}}^{i+1}=\argmin_{\boldsymbol{\alpha},\boldsymbol{c}}r^{i+1}(\boldsymbol{\alpha},\boldsymbol{c})
\end{equation}
where $r^{i+1}(\boldsymbol{\alpha},\boldsymbol{c})$ is the \textit{Rothe error}
\begin{equation}\label{eq:RE_noVarPro}
r^{i+1}(\boldsymbol{\alpha},\boldsymbol{c})=\norm{\sum_{m=1}^{N(t)}c_m\hat A_i \phi_m(\boldsymbol{\alpha})-\hat A_i^\dagger\Psi(t_i)}^2.
\end{equation}
Using the Variable Projection (VarPro) algorithm by Golub and Pereyra, \cite{Golub_Pereyra} the optimal linear coefficients can be obtained from a given set of nonlinear coefficients, and the Rothe error can be rewritten as a function of $\boldsymbol{\alpha}$ only,
\begin{align} \label{eq:RotheError}
    \text{r}^{i+1}(\boldsymbol{\alpha})=\norm{\sum_{m=1}^{N(t)}c_m(\boldsymbol{\alpha})\hat A_i \phi_m(\boldsymbol{\alpha})-\hat A_i^\dagger\Psi(t_i)}^2,
 \end{align}
where 
\begin{equation}\label{eq:solve_lincoeff}
    \boldsymbol{c}^{i+1}(\boldsymbol{\alpha})=\left[ \boldsymbol{S}^{i+1}(\boldsymbol\alpha)\right]^{-1}\boldsymbol{\rho}^{i+1}(\boldsymbol \alpha),
\end{equation}
and we defined 
\begin{align}
    S^{i+1}_{mn}(\boldsymbol{\alpha}) &= \braket{\hat A_i \phi_m(\boldsymbol{\alpha})|\hat A_i \phi_n(\boldsymbol{\alpha})} \nonumber \\
    &= \braket{\phi_m(\boldsymbol{\alpha})|\hat A_i^\dagger \hat A_i|\phi_n(\boldsymbol{\alpha})},\label{eq:Smat} \\ 
    \rho^{i+1}_{m}(\boldsymbol{\alpha}) &= \braket{\hat A_i \phi_m(\boldsymbol{\alpha})|\hat A_i^\dagger\Psi(t_i)} \nonumber \\
    &= \braket{\phi_m(\boldsymbol{\alpha})|\hat A_i^\dagger \hat A_i^\dagger|\Psi(t_i)}.\label{eq:rhovec}
\end{align}

Using the VarPro algorithm simplifies the optimization by reducing the number of variables. As is evident from \cref{eq:Smat,eq:rhovec}, one is required to calculate overlap matrix elements, Hamiltonian matrix elements and squared Hamiltonian matrix elements in order to evaluate the Rothe error.

Writing the resulting Rothe error and the wave function at that time step as 
\begin{align}
    r^{i+1}_{\text{opt.}}&=\min_{\boldsymbol{\alpha},\boldsymbol{c}}r^{i+1}(\boldsymbol{\alpha},\boldsymbol{c})=r^{i+1}(\boldsymbol{\alpha}_{\text{opt.}}^{i+1},\boldsymbol{c}_{\text{opt.}}^{i+1})\\
    \Psi(t_{i+1})&=\sum_{m=1}^{N(t)} \left(c_{\text{opt.}}^{i+1}\right)_m\phi_m(\boldsymbol{\alpha}_{\text{opt.}}^{i+1}),
\end{align}
it can be shown\cite{schrader2024time} that the Rothe error is an upper bound for the Crank-Nicolson time evolution error for hermitian Hamiltonians:
\begin{equation}
        \left\|\Psi(t_{i+1})-(\hat A_{i})^{-1}\hat A_{i}^\dagger\Psi\left(t_i\right)\right\|^2\str{<}\newb{\leq} r^{i+1}_{\text{opt.}}.
\end{equation}
For sufficiently small time steps, i.e., when the Crank-Nicolson propagator is a good approximation to the exact propagator, the cumulative Rothe error
\begin{equation}
    r^\text{cumul.}(t)=\sum_{i=0}^{n_{t}}\sqrt{r^{i+1}_{\text{opt.}}},
\end{equation}
where $n_t$ is the number of time steps going from the
initial time $t_i$ to time $t$, can be used as an approximation to an upper bound of the time evolution error
\begin{equation}
    \text{err}(t)=\norm{\Psi(t)-\Psi_\text{exact}(t)}.
\end{equation}
\subsection{Calculation of matrix elements}
The product of two Explicitly Correlated Gaussians is itself an Explicitly Correlated Gaussian. The effect of the kinetic energy operator acting on an ECG gives a polynomial multiplied by an ECG, and so does an ECG times a polynomial potential $\hat{V}$, such as is the case for the Henon-Heiles potential, see Eq.~\eqref{eq:hamiltonian} in Sec.~\ref{sec:potential-initial}. Furthermore, the derivative of matrix elements with respect to nonlinear parameters becomes integrals of polynomials multiplied with ECGs. Thus, in order to calculate the overlap matrix, the Hamiltonian matrix and the squared Hamiltonian matrix, one only needs to calculate matrix elements of the form 
\begin{equation}\label{eq:Overlap}
    \int \exp(-\boldsymbol r^T\boldsymbol{A}\boldsymbol r + \boldsymbol j^T \boldsymbol r)d\boldsymbol r,
\end{equation}
for complex symmetric $\boldsymbol{A}$ and complex $\boldsymbol{j}$, as well as matrix elements of the form 
\begin{equation}\label{eq:Poly}
    \langle r_ir_j\dots r_n\rangle=\int r_ir_j\dots r_n \exp(-\boldsymbol r^T\boldsymbol{A}\boldsymbol r + \boldsymbol j^T \boldsymbol r)d\boldsymbol r,
\end{equation}
where $r_i,r_j,\dots,r_n\in [r_1,\dots,r_D]$. Equation \eqref{eq:Overlap} has a well known result,
\begin{align}\label{eq:Overlap_sol}
    \int \exp(-\boldsymbol r^T\boldsymbol{A}\boldsymbol r + \boldsymbol j^T \boldsymbol r)d\boldsymbol r &=\frac{\pi^{D/2}}{\det(\sqrt{\boldsymbol A})} \nonumber \\
    &\times \exp\left(\frac{1}{4}\boldsymbol{j}^T \boldsymbol{A}^{-1} \boldsymbol{j}\right),
\end{align}
where $\sqrt{\boldsymbol{A}}$ is the principal square root of $\boldsymbol{A}$. If $\boldsymbol{A}$ is diagonalizable, we have that
\begin{equation}
    {\det(\sqrt{\boldsymbol A})}=\prod_{i=1}^D \sqrt{\lambda_i},
\end{equation}
where $\sqrt{\lambda_i}$ is the principal square root of the i'th eigenvalue $\lambda_i$ of $\boldsymbol{A}$. Observe that
\begin{equation}
    {\det(\sqrt{\boldsymbol A})}\neq {\sqrt{\det({\boldsymbol A})}},
\end{equation}
contrary to what is sometimes found in the literature.
Polynomial expectation values can be calculated using Isserlis' theorem (also known as Wick's probability theorem). \cite{Isserlis_1918,Wick_1950,Withers_1985} For a complex multivariate normal distribution 
\begin{equation}
    p(\boldsymbol{r})=\frac{\operatorname{det}\left(\sqrt {\boldsymbol{A}}\right)}{\pi^{D / 2}}  \exp \left(-(\boldsymbol{r}-\boldsymbol{\mu})^{\mathrm{T}} \boldsymbol{A}(\boldsymbol{r}-\boldsymbol{\mu})\right),
\end{equation}
and defining $\boldsymbol{x}=\boldsymbol{r}-\boldsymbol{\mu}$, one can calculate expectation values as 
\begin{align}\label{eq:Isserlis}
        \langle x_ix_j\dots x_n\rangle&=\int p(r) x_ix_j\dots x_ndr\nonumber \\&=\frac{1}{2}\sum_{p \in P_n^2} \prod_{\{i, j\} \in p}\left(\boldsymbol{A}^{-1}\right)_{ij},
\end{align}
where the sum is taken over all unique pair combinations of the $n$ variables (which is zero if $n$ is odd), and the product is over the $n/2$ pairs within each combination. As the integrand in eq. \eqref{eq:Poly} can be rewritten as a sum over integrals of the form of eq. \eqref{eq:Isserlis}, there exist analytical formulas to calculate all matrix elements of interest.\\

\subsection{Optimization of nonlinear parameters}\label{sec:Optimization}
The Rothe error, \cref{eq:RotheError}, is a differentiable function of the nonlinear parameters $\boldsymbol{\alpha}$.
In this paper, we use the Broyden-Fletcher-Goldfarb-Shanno (BFGS) algorithm~\cite{nocedal_numerical} implemented in SciPy~\cite{scipy} to carry out the optimization using analytical gradients. As initial guess for the nonlinear coefficients at time step $t_{i+1}$, we used the initial coefficients at the previous time step $t_{i}$, and whenever applicable, also added a fraction $\delta$ of the change going from ${t_{i-1}}$ to $t_i$, i.e.
\begin{equation}
\boldsymbol{\alpha}_{\text{init.}}^{i+1}=\boldsymbol{\alpha}_{\text{opt.}}^{i}+\delta(\boldsymbol{\alpha}_{\text{opt.}}^{i}-\boldsymbol{\alpha}_{\text{opt.}}^{i-1}),
\end{equation}
where the optimal value for $\delta\in[0,1]$ was found using a line search to find the value of $\delta$ that led to the lowest Rothe error. \\
If an unconstrained optimization is carried out, parameters of some Gaussians change in such a way in the optimization procedure that the Gaussians might no longer contribute. In that case, one might end up in a situation where Gaussians become very hard to re-introduce into the wave functions, as their parameters need to change significantly. This problem is reminiscent of the vanishing gradient problem in Machine Learning \cite{Vanish_gradient1,Vanish_gradient2} and the barren plateau problem in Quantum Machine Learning. \cite{BarrenPlateau1,BarrenPlateau2} In order to circumvent this problem, optimization was carried out with a transformed set of parameters $(\boldsymbol{\alpha}^{i+1})'$ instead,
\begin{align}
    (\boldsymbol{\alpha}^{i+1})_j  &= {{\min}_j} \nonumber \\
    &+ \big(\tanh( (\boldsymbol{\alpha}^{i+1})'_j)\! +\! 1\big) \frac{({\rm {\max}_j} - { {\min}_j})}{2},
 \end{align}
 where
\begin{align}
{\min}_j&=\boldsymbol{(\alpha}^{i+1}_{\text{init.}})_j-p|\boldsymbol{(\alpha}^{i+1}_{\text{init.}})_j|-q,\\
{\max}_j&=\boldsymbol{(\alpha}^{i+1}_{\text{init.}})_j+p|\boldsymbol{(\alpha}^{i+1}_{\text{init.}})_j|+q.
\end{align}
We have chosen $p=3$ and $q=0.5$ for all calculations considered. As an initial guess for the the Hessian matrix at the first iteration $H_0$, we used the diagonal matrix containing the elements of the absolute value of the gradient of the initial Rothe error
\begin{equation} 
H_0=\text{diag}\left(|{\nabla_{\boldsymbol\alpha}r_{i+1}\left(\boldsymbol{\alpha}^{\text{init.}}_{i+1}\right)}|\right),\end{equation}
which ensures that the optimization is scale invariant.
\subsection{Masking functions}
\begin{figure}[h!]
    \centering
    \includegraphics[width=\linewidth]{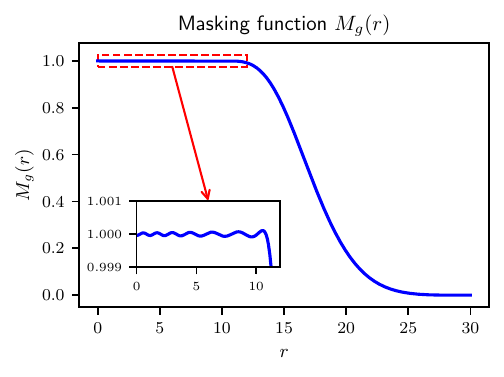}
    \caption{\justifying Gaussian masking function $M_g(r)$ consisting of $n_{\text{mask}}=20$ Gaussians, with $r_0=11$ and $r_1=25$.}
    \label{fig:MaskFunction}
\end{figure}

As the Henon-Heiles potential is unbounded, parts of the wave function will escape from the potential well. The escaping part of the wave function will have a very large local energy, which will result in large Rothe errors, and simulating it is not of interest. Therefore, special care has to be taken of outgoing wave packets. While our initial idea was to remove Gaussians whose centers are further than 12 units away from the origin, as was done in Ref. \citenum{Burghardt_Worth_HeHe}, we found that this approach was insufficient. Even though the center of an ECG can be far away from the origin, due to its varying width and oscillating character and interference with the other ECGs present, it can still contribute substantially to the wave function also in the regions of space that we are interested in. Hence, simple removal can lead to unpredictable artifacts. Thus, we adopted a different approach. While complex absorbing potentials \cite{Manolopoulos,Giovannini,Yu_2018} and masking functions \cite{Masking_functions} on a grid can be modeled in a straightforward way, as one only needs to calculate their action at every grid point, their application to wave functions written as linear combinations of ECGs become considerably more involved. The effect of a masking function ${M}(\boldsymbol{r})$ is to replace the time-dependent wave function $\Psi(\boldsymbol{r},t)$  by $ {M}(\boldsymbol{r})\Psi(\boldsymbol{r},t)$ at every time step. The masking function is chosen such that the relevant dynamics are preserved by keeping the relevant part of the wave function while the irrelevant parts are removed. A popular choice is the radial $\cos^{1/8}$ masking function,
\begin{equation}
  M_c(r) =
    \begin{cases}
      1 & \text{if $r\leq r_0$}\\
      \cos^{1/8}\left(\frac{\pi}{2}\frac{r-r_0}{r_1-r_0}\right) & \text{if $r_0<r<r_1$}\\
      0 & \text{if $r\geq r_1$}
    \end{cases}.
\end{equation}
However, to our knowledge, no closed form solutions for matrix elements involving products of EGCS and $M_{c}(r)$ exist. Hence, we define a masking function written as linear combination of Gaussians,
\begin{equation}
    M_g(r)=\sum_{i=1}^{n_{\text{mask}}}d_ie^{-\gamma_i r^2}
\end{equation}
where the coefficient vectors $\boldsymbol{d}$ and $\boldsymbol{\gamma}$ are chosen in such a way that $M_g(r)\approx 1$ for $r<r_0$ and $M_g(r)\approx 0$ for $r>r_1$. Setting $r_0=11$, $r_1=25$ and $n_{\text{mask}}=20$, the coefficient vectors $\boldsymbol{d}$ and $\boldsymbol{\gamma}$ were found on a 1D grid using numerical optimization and the Variable Projection algorithm. \cite{Golub_Pereyra,OLeary2013} The resulting masking function is shown in figure \ref{fig:MaskFunction}.

However, a drawback of this masking function is that the number of ECGs would be increased by a factor of $n_{\text{mask}}$ \emph{at every time step}. In order to avoid this, we refitted ${M}_g({r})\Psi(\boldsymbol{r},t)$ with $N(t)$ Gaussians, where $N(t)$ is the number of ECGs in $\Psi(\boldsymbol{r},t)$ at time step $T$. As the initial guess, we simply used $\Psi(\boldsymbol{r},t)$.
To further reduce the cost of using the masking function, this procedure is only applied every 5\textsuperscript{th} time step, and only if 
\begin{equation}
    ||{M}_g(\boldsymbol{r})\Psi(\boldsymbol{r},t)-\Psi(\boldsymbol{r},t)||>\varepsilon_{\text{mask}},
\end{equation}
where we set $\varepsilon_{\text{mask}}= 10^{-3}$. Observe also that $||{M}_g(\boldsymbol{r})\Psi(\boldsymbol{r},t)-\Psi(\boldsymbol{r},t)||$ can be of the magnitude $\sim 10^{-5}$ even if $\Psi(\boldsymbol{r},t)$ essentially does not extend beyond $r_0$. This is because the masking function is not exactly identical to $1$ even for $r\in[0,r_0]$ as it is written as a linear combination of Gaussians. This gives a lower limit for sensible choices of $\varepsilon_{\text{mask}}$. There is, however, no lower limit for $\varepsilon_{\text{mask}}$, as the accuracy can be improved simply by increasing the number of Gaussians $n_{\text{mask}}$.
As there is no underlying grid, the fact that ECGs might move beyond $r_0$ and even beyond $r_1$ between successive applications of the masking procedure does not cause any issues like unphysical reflections. 

\subsection{Conservation of norm and energy}\label{sec:Conservation}

The time-dependent Schrödinger equation conserves the norm of the wave function, and for time-independent Hamiltonians, the energy is also conserved. While the Crank-Nicolson method is symplectic and thereby conserves both norm and energy, Rothe's method only approximates the Crank-Nicolson method, and as such, neither norm nor energy are strictly conserved. \newb{The Rothe error is a global measure of the deviation from exact Crank-Nicolson propagation, and it is agnostic to where or how it arises. It is hence possible that the minimal Rothe error can be found for a parameterization where norm and energy are not conserved, independent of what parameters are allowed to change, i.e. even if only the linear parameters are allowed to change.} To address this, one can add Lagrange multipliers to eq. \eqref{eq:RotheError} or use an augmented Lagrangian method or a penalty method \cite{nocedal_numerical} to carry out a constrained optimization subject to (approximate) norm and energy conservation. However, the VarPro algorithm will then not be applicable, as there will not be a closed-form solution for the linear parameters. \cite{Sima_vapro}
To circumvent this problem, we first carry out an unconstrained optimization of the Rothe error using the VarPro algorithm, followed by a constrained optimization of the linear parameters subject to conservation of norm and energy. That is, we use the SLSQP algorithm \cite{nocedal_numerical} to carry out the optimization
\begin{equation}
\begin{aligned}
    \boldsymbol{c}_{i+1} = \argmin_{\boldsymbol{c}} \;  r_{i+1}(\boldsymbol{\alpha}_{i+1}^{\text{opt}}, \boldsymbol{c}), & \\
    \text{subject to}\ \left\{ 
    \begin{aligned}
        &\langle\Psi(t_{i+1})|\Psi(t_{i+1})\rangle - n = 0, \\
        &\langle\Psi(t_{i+1})|\hat{H}|\Psi(t_{i+1})\rangle - E = 0,
    \end{aligned}
    \right.
\end{aligned}
\end{equation}
where $n$ is the norm of the wave function and $E$ is the energy. The norm $n$ will not necessarily be constantly equal to $n=1$ as an application of the masking function will change the norm, and similarly, the energy is not conserved when the masking function is applied.
This procedure is efficient, as no new matrix elements need to be calculated. In principle, this procedure does not lead to the optimal norm-conserving and energy-conserving basis, as the nonlinear parameters are not optimized simultaneously. However, we observe that the resulting increase in the Rothe error due to conservation of norm and energy is orders of magnitude smaller than the Rothe error itself for sufficiently many basis functions, which justifies this procedure. This procedure is only applicable if the number of Gaussians is at least $N=2$, as conserving both norm and energy with just one free parameter is not doable. However, instabilities can be encountered for small $N$ as energy conservation requires the basis to be flexible enough for the energy $E$ to be obtainable while still conserving norm. 
\subsection{{Variational dynamics of a single ECG}}\label{sec:singleGaussianDynamics}
A useful feature of the Heller form is that there are closed-form variational solutions for a single Gaussian. \cite{Faou_Lubich_2006,lubich2008quantum} Using Strang splitting, \cite{Strang,McLachlan_Quispel_2002} an approximation to the exact time evolution operator $U(t_i+\Delta t, t_i)$ from time $t_i$ to time $t_i+\Delta t$ reads
\begin{align}\label{eq:StrangSplitting}
    U(t_i+\Delta t, t_i) &= e^{-\frac{i\Delta t}{2}\hat T}e^{-{i \Delta t}\hat{V}(t_i+\frac{\Delta t}{2})}e^{-\frac{i\Delta t}{2} \hat T} \nonumber \\
    &+ O((\Delta t)^3).
\end{align}

When considering the action of $e^{-{i \Delta t}\hat{V}}$ on a single ECG, the time-dependent variational principle applied to an ECG in the Heller form yield a set of differential equations that have analytical solutions, and the same holds for $e^{-\frac{i\Delta t}{2} \hat T}$.
In particular, \citeauthor{lubich2008quantum}~\cite{lubich2008quantum} gives the following equations for the time evolution of the coefficients $\boldsymbol{C},\boldsymbol{q},\boldsymbol{p},\zeta$ in Heller form going from time $t$ to time $t+\Delta t$:
\begin{equation}\label{eq:ExactPropagationKinetic}
    \begin{aligned}
        \boldsymbol{q}(t+\Delta t)&=\boldsymbol{q}(t)+\Delta t \boldsymbol{p}(t),\\
        \boldsymbol{p}(t+\Delta t)&=\boldsymbol{p}(t),\\
        \boldsymbol C(t+\Delta t) &=\boldsymbol C(t)\left(\boldsymbol I+\Delta t \boldsymbol C(t)\right)^{-1}, \\
        \zeta(t+\Delta t) & =\zeta(t)+\frac{\Delta t}{2}\left|\boldsymbol p(t)\right|^2\\
        &+\frac{\text i}{2} \operatorname{tr}\left(\log \left(\boldsymbol I+\Delta t \boldsymbol C(t)\right)\right).
    \end{aligned}
\end{equation}
Furthermore, if $\hat{V}$ is a polynomial of at most second order,
\begin{equation}
    \hat{V}=\sum_i^D c_i r_i+\sum_{i}^D\sum_{j}^D c_{ij} r_ir_j,
\end{equation}
then it can be shown \cite{lubich2008quantum} that a single Gaussian remains Gaussian under time evolution, i.e., there exists a set of parameters $\boldsymbol C(t),\boldsymbol p(t),\boldsymbol q(t),\zeta(t)$ such that $\tilde g(\boldsymbol C(t),\boldsymbol p(t),\boldsymbol q(t),\zeta(t))$ is a solution to the time-dependent Schrödinger equation at all $t$:
\begin{equation}
    \hat{H}\tilde g(\boldsymbol C(t),\boldsymbol p(t),\boldsymbol q(t),\zeta(t))=\text{i}\frac{\partial}{\partial t}\tilde g(\boldsymbol C(t),\boldsymbol p(t),\boldsymbol q(t),\zeta(t)).
\end{equation}
Hence, in the limit $t\rightarrow0$, using eq. $\eqref{eq:StrangSplitting}$ and solving the variational equations of motion, the exact solution can be found.
\subsection{Application of Strang splitting to Rothe's method with Gaussians}\label{sec:StrangSplitting}
By the discussion in the previous subsection, for any normalizable Gaussian $\tilde g(\boldsymbol C,\boldsymbol p,\boldsymbol q,\zeta)$, acting with the kinetic energy propagator $e^{-\frac{i\Delta t}{2} \hat T}$ yields a new Gaussian $\tilde g(\boldsymbol C',\boldsymbol p',\boldsymbol q',\zeta')$. Specifically, by eq. \eqref{eq:ExactPropagationKinetic},
\begin{equation}
e^{-\frac{i\Delta t}{2} \hat T}\tilde g(\boldsymbol C,\boldsymbol p,\boldsymbol q,\zeta)=\tilde g(\boldsymbol C',\boldsymbol p',\boldsymbol q',\zeta')=\tilde{g}'.
\end{equation}
By linearity of the kinetic energy propagator, a linear combination of Gaussians can thus be propagated independently without introducing any approximation. By first applying Strang splitting and then replacing the potential energy propagator with the Crank-Nicolson propagator, the full propagator can be approximated as 
\begin{equation}
    U(t_i+\Delta t, t_i) =
    e^{-\frac{i\Delta t}{2}{\hat{T}}}\hat{B}_i^{-1}\hat{B}_i^\dagger e^{-\frac{i\Delta t}{2}{\hat{T}}}+O((\Delta t)^3),
\end{equation}
where 
\begin{equation}
    \hat{B}_i=\left({\hat{I}+\text{i}\frac{\Delta t}{2}\hat{V}\left(t_i+\frac{\Delta t}{2}\right)}\right).
\end{equation}
One can now separate kinetic and potential propagation in Rothe's method: First, the wave function, written as a linear combination of Gaussians, is propagated with the kinetic energy propagator only with time step $\Delta t/2$. Then, Rothe's method is used to carry out the propagation stemming from the potential with time step $\Delta t/2$ to approximately propagate the Gaussians inter-dependently. Finally, a propagation with time step $\Delta t/2$ is carried out with the kinetic energy operator. Doing this, one no longer needs to evaluate the full squared Hamiltonian $\hat{H}^2$, only the squared potential $\hat{V}^2$ is required.
The Strang-splitting approach could be further exploited by using Rothe's method only for the anharmonic part of the potential, but we have not pursued this idea in the present work.
We did implement an energy- and norm-conservation scheme similar to the one introduced in Section \ref{sec:Conservation}, differing only in that not the energy $\langle\hat{H}\rangle$, but the expectation value of the potential $\langle\hat{V}\rangle$ should be conserved in the Rothe step. As Strang splitting is not energy-conserving, however, this does not guarantee full energy conservation.
\section{Model system and implementation}\label{sec:system-implementation}
\subsection{The Henon-Heiles potential and initial conditions}
\label{sec:potential-initial}
The Hamiltonian for the $D$-dimensional Henon-Heiles potential reads
\begin{align}
    \hat{H} &=-\frac{1}{2} \sum_{i=1}^D \frac{\partial^2}{\partial r_i^2}+\frac{1}{2} \sum_{i=1}^D r_i^2 \nonumber \\
    &+\lambda \sum_{i=1}^{D-1}\left(r_i^2 r_{i+1}-\frac{1}{3} r_{i+1}^3\right), \label{eq:hamiltonian}
\end{align}
where the first sum represents the kinetic energy, the second sum corresponds to $D$ one-dimensional harmonic oscillators, and the third sum is an anharmonic term that gives raise to complicated dynamics. We set the strength of the anharmonicity $\lambda=0.111803\approx 1/\sqrt{80}$, which was also used in Refs.~\citenum{Burghardt_Worth_HeHe} and \citenum{NestHeHe}. The Henon-Heiles potential is a test potential that does not represent any specific physical system. By adjusting the strength of the harmonic term and the mass of the particle, several natural unit systems can yield this Hamiltonian (\cref{eq:hamiltonian}). Therefore, we report energies and times without units, as any specific choice would be arbitrary.

The initial state is a single normalized Gaussian with standard width displaced from the origin by $2$ in all dimensions. That is,
\begin{equation}\label{eq:InitialGaussian}
\Psi(\boldsymbol{r},t=0)=\pi^{-D/4} \exp\left(-\left(\boldsymbol{r}-\boldsymbol{\mu}^0\right)^T\boldsymbol{A}^0\left(\boldsymbol{r}-\boldsymbol{\mu}^0\right)\right),
\end{equation}
where $\boldsymbol{A}^0=\frac{1}{2}\boldsymbol{I}_D$ and $\mu^0_i=2$ for $i=1,\dots,D$.
\newb{With these initial conditions, the resulting dynamics deviate significantly from those observed in a purely harmonic potential, where the wave function could be represented with a single ECG at all times. This behavior is illustrated in figure \ref{fig:2D_example}, which shows the wave function at various time steps as it evolves under the dynamics of the Henon-Heiles potential.
\begin{figure}
    \centering
  \includegraphics[width=0.5\textwidth]{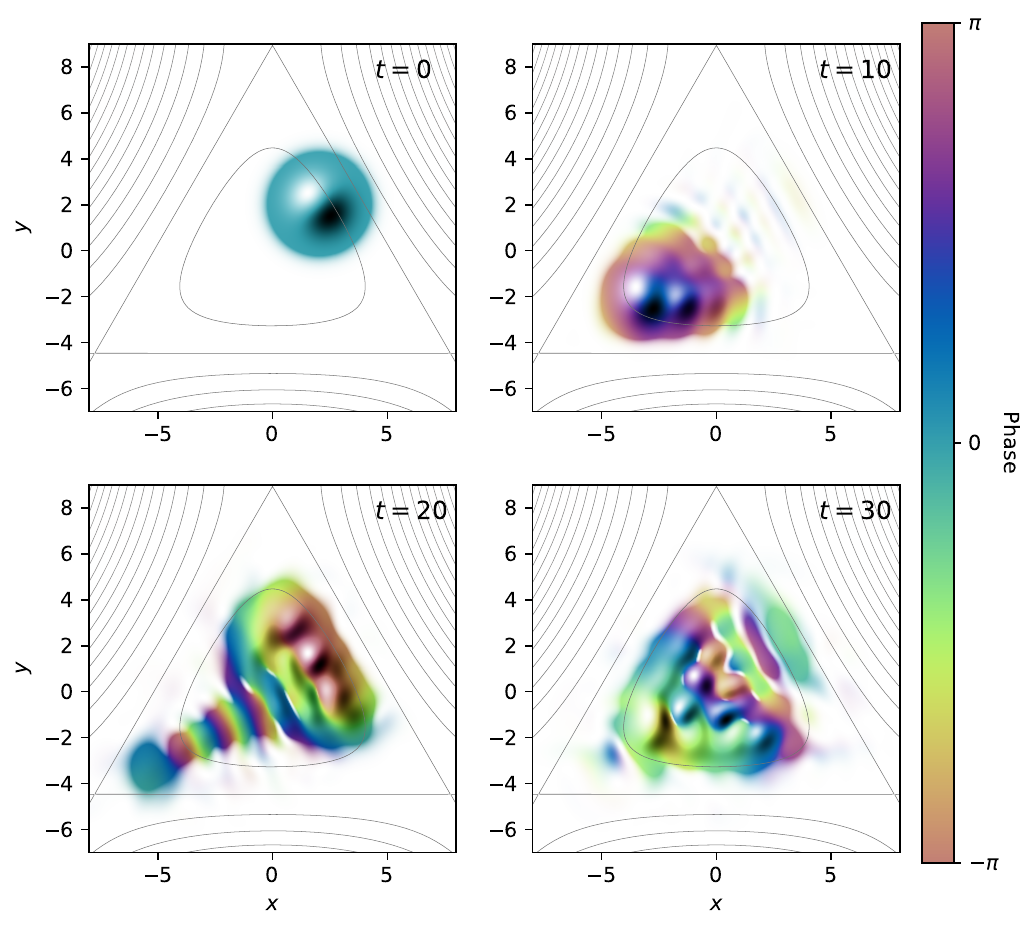}
  \caption{Illustration of the time evolution of the initial state (eq. \eqref{eq:InitialGaussian}) in the Henon-Heiles potential, illustrating the wave function at $t=0$, $t=10$, $t=20$ and $t=30$. The color represents the phase, while the apparent height represents the amplitude.}
  \label{fig:2D_example}
\end{figure}
}
\subsection{Spectrum and fidelity}
Following Ref. \citenum{NestHeHe}, we consider the spectrum $S(\omega)$, which is given as the real part of the inverse Fourier transform of the autocorrelation function $C(t)$
\begin{equation}
    \label{eq:spectrum}
    S(\omega)\propto \operatorname{Re} \int_0^{\infty} \exp\left(i \omega t-t/30\right) C(t) d t,
\end{equation}
where $\exp\left(-t/30\right)$ serves as a damping factor.
The autocorrelation function $C(t)$ is given as
\begin{equation}
    C(t)=\langle \Psi(0)|\Psi(t)\rangle.
\end{equation}
For real initial states $\Psi(0)=\Re\left(\Psi(0)\right)$, this is formally identical to 
\begin{equation}
    C(t)=\langle \Psi^*(t/2)|\Psi(t/2)\rangle,
\end{equation}
which is used in our calculations.

In order to evaluate the convergence of the method with increasing number of Gaussians, we also consider the fidelity (which for pure states corresponds to the transition probability). \cite{nielsenchuang} The fidelity between two (normalized) quantum states $\ket{\Psi_1}$ and $\ket{\Psi_2}$ is defined as 
\begin{equation}
F(\Psi_1,\Psi_2)=|\!\braket{\Psi_1|\Psi_2}\!|^2,
\end{equation}
and is a measure for the degree of similarity between the two states. For simulations with the same conditions (i.e., the dimension $D$ and the same time step $\Delta t$), but with different numbers of Gaussians used, we consider the time-dependent fidelity 
\begin{equation*}
    F^D_{N,N_{\text{max}}}(t)=F\left(\Psi^{N_{\text{max}}}(t),\Psi^{N}(t)\right),
\end{equation*}
where $\Psi^{N}(t)$ is the wave function consisting of $N$ Gaussians at time $t$, $D$ the dimension, and $N_{\text{max}}$ the maximal number of Gaussians considered. With increasing number $N$ of Gaussians, the fidelity $F^D_{N,N_{\text{max}}}(t)$ is expected to increase for all $t$, as this is an indication that the dynamics converge to the same solution. It can also be seen as a measure of convergence to the solution---if the fidelity at $t=T$ is large when $N<N_{\text{max}}$, it means that increasing the number of Gaussians no longer leads to a very different solution.

\subsection{Comparison schemes}
In order to benchmark our simulation, we compare with grid simulations. We used a Cartesian grid of size $512\times512$ ($2D$), $128\times128\times128$ ($3D$) and $64\times64\times64\times64$ ($4D$) with a spatial extent of $r_i \in [-25,25]$ for each each dimension. We used a split-step Fourier scheme to propagate the wave function on the grid~\cite{feit_solution_1982} with time step $\Delta t=0.01$. On the grid, we used a $\cos^{1/8}$ masking function with $r_0=11$, $r_1=25$. Because different integration schemes were used for Rothe's method and the grid, and because the action of the mask is slightly different, we do not expect exact quantitative agreement between wave functions and expectation values even when the Rothe calculation has fully converged with respect to number of Gaussians, but we do expect the results to be very similar. 
\subsection{Computational details}
In all computations, we use normalized basis functions in order to avoid problems with numerical accuracy. 
We calculated the wave function up to time $T=100$, thereby obtaining the autocorrelation function up to time $2T=200$. We used the time step size $\Delta t=0.01$ in all simulations, thus, the number of time steps is  $n_T=10,000$. This time step was chosen by considering the dynamics of a single Gaussian in a $2D$ harmonic oscillator potential. Using a time step $\Delta t =0.01$, the cumulative Rothe error at $T=100$ was $0.02$, which indicates essentially exact dynamics, while using $\Delta t=0.025$, the cumulative Rothe error was $0.5$, which indicates a too large time step. While the Rothe scheme is inherently adaptive and allows for a variable number of Gaussians, which can be increased---e.g., whenever the Rothe error gets bigger than a threshold $\varepsilon_{\text{max}}$\cite{kvaal2023need,schrader2024time}---we did not use this adaptivity here. Instead, we started the calculation with a fixed number $N$ of Gaussians, using the initial state (\cref{eq:InitialGaussian}) and a number of $N-1$ Gaussians with the same initial correlation matrix $\boldsymbol{A}^0=\frac{1}{2}\boldsymbol{I}_D$, placed in the vicinity of the initial state.
Specifically, we set the initial position of each Gaussian at $t=0$, $\boldsymbol{\mu}^0_m$, as
\begin{equation}
\boldsymbol{\mu}^0_m=\boldsymbol{\mu}^0+\boldsymbol{x}_m
\end{equation}
for $m=2,\dots,N$ (with $\boldsymbol{\mu}^0_1=\boldsymbol{\mu}^0$), where $\boldsymbol{x}_m$ is sampled from a uniform distribution on $[-0.25,0.25]^D$. \newb{This choice ensures that the Gaussians may be populated from the first time step also with the constrained optimization described in section \ref{sec:Optimization}, while avoiding numerical issues.} In the discussion, we will come back to arguing why adaptivity was not used. 

Normalized Gaussians where removed when the associated linear coefficient is less than $10^{-6}$ times the largest linear coefficient, i.e., Gaussian $i$ is removed when
\begin{equation}
    |c_i|<10^{-6}\max{|\boldsymbol{c}|}.
\end{equation}
This mainly happens when a Gaussian basis function moves far away from the potential well near the origin. In that case, it will have a very small linear coefficient due to the masking function and, therefore, should be removed.

To avoid numerical issues when Gaussians are very close and the overlap matrix is near-singular, which mainly happens at the beginning of the propagation and in $2D$ due to the initialization of the system, a small regularization parameter $\lambda$ was added to the inverse when solving for the linear coefficients in eq. \eqref{eq:solve_lincoeff}, i.e., the linear coefficients are instead obtained as 
\begin{equation}
    \boldsymbol{c}_{i+1}(\boldsymbol{\alpha})=\left[ \boldsymbol{S}^{i+1}(\boldsymbol\alpha)+\lambda\boldsymbol{I}\right]^{-1}\boldsymbol{\rho}^{i+1}(\boldsymbol \alpha),
\end{equation}
 where $\lambda=10^{-8}$. 

The code for this program is written in Python, and we made use of mpi4py \cite{mpi4py_2005,mpi4py_2008,mpi4py_2011,mpi4py_2021} to parallelize the calculation of the matrix elements and their derivatives.
\subsection{\protect\newb{Computational scaling}}
\newb{The computational cost is dominated by the calculation of matrix elements involving the squared Hamiltonian, $\hat{H}_{mn}^2(\boldsymbol{\alpha})=\langle g_m(\boldsymbol{\alpha})|\hat H^2|g_n(\boldsymbol{\alpha})\rangle$ and the calculation of their derivatives with respect to the nonlinear coefficients
\begin{equation}
    \frac{\partial\hat{H}_{mn}^2(\boldsymbol{\alpha})}{\partial \boldsymbol \alpha}.
\end{equation}
For Hamiltonians where the potential consists of polynomial terms only, such as the Henon-Heiles Hamiltonian (eq. \eqref{eq:hamiltonian}), all matrix elements are sums over expectation values of polynomial terms, which can be calculated with Isserlis' theorem (eq. \eqref{eq:Isserlis}).
In particular, when the potential has polynomial terms of order up to $P$, the squared Hamiltonian has polynomial terms of order up to $2P$. The derivative of matrix elements with respect to the $\textbf{A}_m$-matrix, i.e. 
\begin{equation}\label{eq:Aderiv}
    \frac{\partial\hat{H}_{mn}^2(\boldsymbol{\alpha})}{\partial \boldsymbol A_m}.
\end{equation}
requires calculating expectation values of polynomial terms of order $2P+2$. For a general Hamiltonian, we can assume that there are $O(D^k)$ such terms - for the Henon-Heiles potential, there is just a single sum over third-order polynomials, so $k=1$, and in general, $k\leq P$. In the squared Hamiltonian, there are hence $O(D^{2k})$ polynomial terms of order $2P$, and taking into account the derivatives, there are $O(D^{2k+2})$ terms of order $2P+2$. The number of all unique pair combinations required in Isserlis' theorem, scales double-factorially in $P-1$ (however, it should be underlined that $P$ generally does not grow with $D$), and the amount of matrix elements, and hence derivatives required, is $N(N+1)/2=O(N^2)$ for $N$ Gaussians. In addition, we have $n_{iter}$ amount of iterations in the optimization procedure for each time step, and $n_T$ time steps in total.
To conclude, the overall scaling is
\begin{equation}\label{eq:scaling}
O(D^{2k+2}(2P+1)!!N^2n_{iter}n_T).
\end{equation}
where, for the Henon-Heiles potential, $k=1$ and $P=3$. We have observed that $n_{iter}$ is roughly proportional to the number of parameters, i.e. the optimization takes a few dozen iterations with $10$ Gaussians in $2D$, but over a hundred iterations with $40$ Gaussians in $4D$. \\ 
It should be mentioned that this analysis has not taken into account that many expectation values are identical (e.g. $\langle x_1x_2\rangle=\langle x_2x_1\rangle$). For example, there at most
\begin{equation*}
    \binom{D+P-1}{D-1}
\end{equation*}
distinct polynomials in a $D$-dimensional Hamiltonian with the largest polynomial length $P$, and this number is much smaller than $D^P$. Furthermore, the possibility of using intermediates or neglecting matrix elements between sufficiently distant Gaussians (which don't need to be calculated), is not taken into account.
Nevertheless,  this scaling is steep - for high-dimensional systems, viable alternatives might involve approaches where the correlation matrices are not dense, but rather \mbox{(block-)diagonal}. This avoids the double-factorial scaling that arises from Isserli's theorem. For diagonal correlation matrices, in particular, expectation values of arbitrary polynomial strings become simple products over single-Gaussian expectation values. Alternatively, if a potential $V(\boldsymbol{x})$ can effectively be expressed as a linear combination of Gaussians, and that potential squared $V^2(\boldsymbol{x})$ can be expressed using a different linear combination of Gaussians (as is the case for the Coulomb potential \cite{schrader2024time}), Isserli's theorem can be avoided as matrix elements reduce to (derivatives of) eq. \eqref{eq:Overlap_sol}.}
\section{Results}\label{sec:Results}
\subsection{Spectra}
\begin{figure}[h!]
    \centering
    \includegraphics[width=\linewidth]{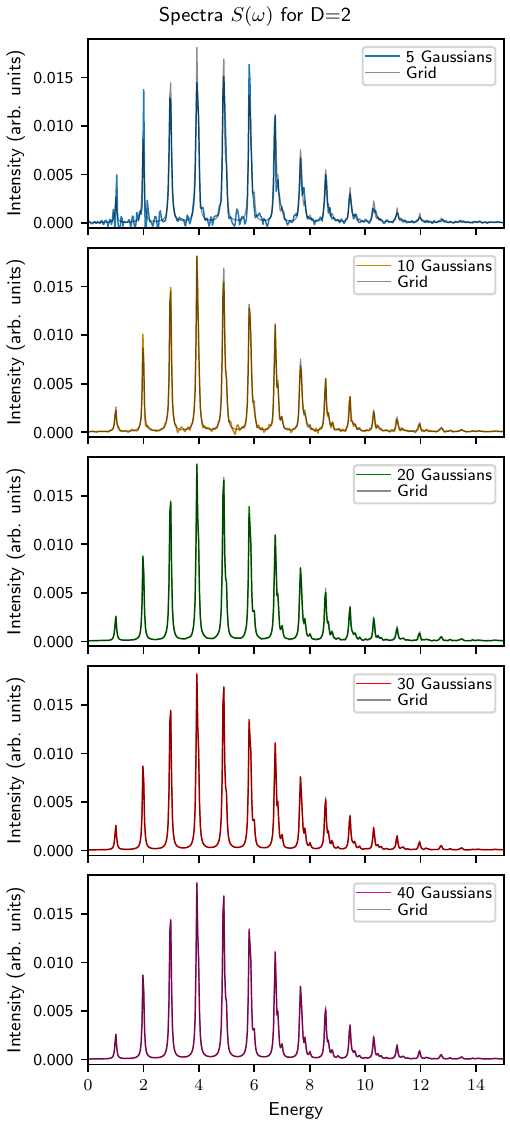}
    \caption{\justifying Spectra for the $2D$ Henon-Heiles potential obtained using $5$, $10$, $20$, $30$, and $40$ Gaussians, compared to the grid reference.}
    \label{fig:2D_spectra}
\end{figure}
\begin{figure}[h!]
    \centering
    \includegraphics[width=\linewidth]{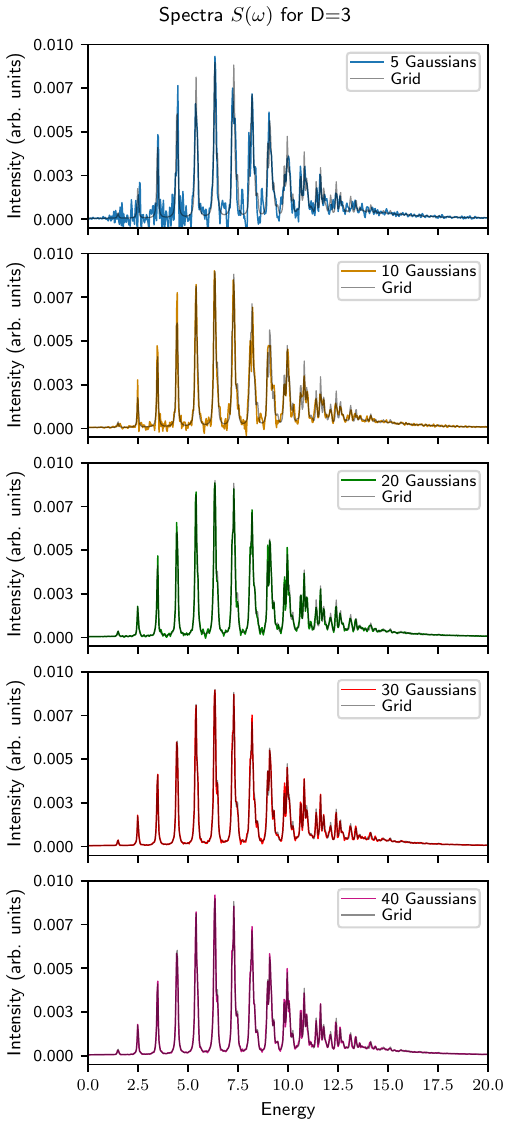}
    \caption{\justifying Spectra for the $3D$ Henon-Heiles potential obtained using $5$, $10$, $20$, $30$, and $40$ Gaussians, compared to the grid reference.}
    \label{fig:3D_spectra}
\end{figure}

Figure \ref{fig:2D_spectra} 
shows the spectra obtained for the $2D$ Henon-Heiles potential using $5$, $10$, $20$, $30$, and $40$ Gaussians, compared to the grid reference. One clearly sees a big improvement in the spectra going from $5$ to $10$ Gaussians, with the peak heights being approximately on grid-level quality, and from $10$ to $20$ Gaussians, with the spectrum for $20$ Gaussians being essentially noise-free.

Figure \ref{fig:3D_spectra} shows the spectra obtained for the $3D$ Henon-Heiles potential using $5$, $10$, $20$, $30$, and $40$ Gaussians, compared with the grid reference. As in the $2D$ case, a clear improvement of the spectrum with increasing number of Gaussians is observed, with the spectra obtained using $20$ and $30$ Gaussians having very little noise and reproducing the peak structure quantitatively. Nearly all noise is gone using $40$ Gaussians, even though not to the same extent as $30$ or $40$ Gaussians in $2D$.

\begin{figure}[h!]
    \centering
    \includegraphics[width=\linewidth]{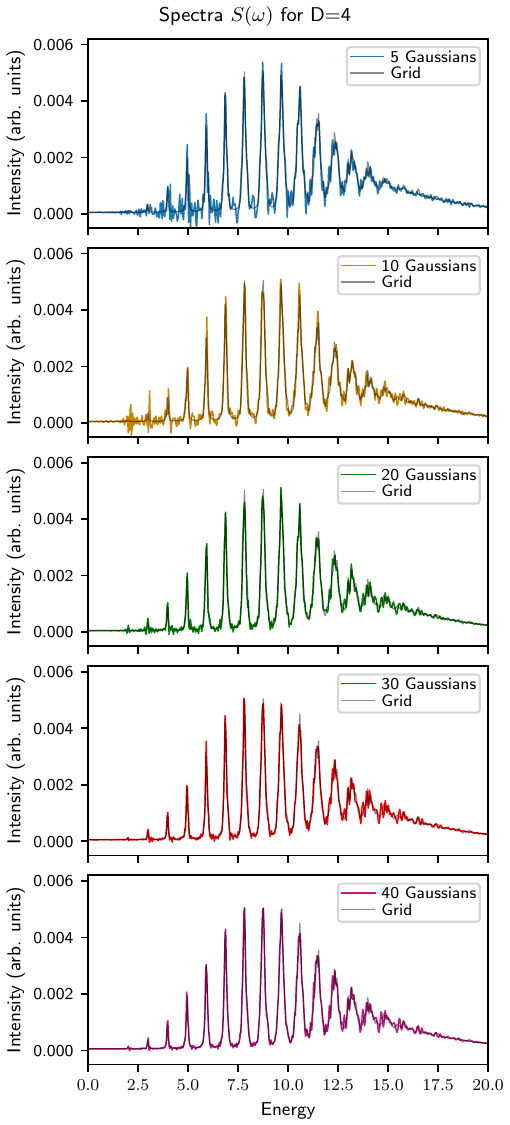}
    \caption{\justifying Spectra for the $4D$ Henon-Heiles potential obtained using $5$, $10$, $20$, $30$, and $40$ Gaussians, compared to the grid reference.}
    \label{fig:4D_spectra}
\end{figure}
The spectra for the $4D$ Henon-Heiles potential using $5$, $10$, $20$, $30$, and $40$ Gaussians, compared to the grid reference, are shown in figure \ref{fig:4D_spectra}. In the $4D$ spectrum using $5$ Gaussians, we turned off the energy conservation at $t=70$ as the dynamics became numerically unstable---this is likely not due to the energy conservation itself, but to the simplified conservation algorithm. With an increased number of Gaussians, we observe strongly improved spectra, but we do not observe a convergence of the spectra, as even the $N=40$ spectrum is relatively noisy, even though the noise is clearly reduced, especially at low energies, as the number of Gaussians is increased to $N=40$. 

\subsection{Fidelities, cumulative Rothe errors and autocorrelation functions}

\begin{figure}[h!]
    \centering
    \includegraphics[width=\linewidth]{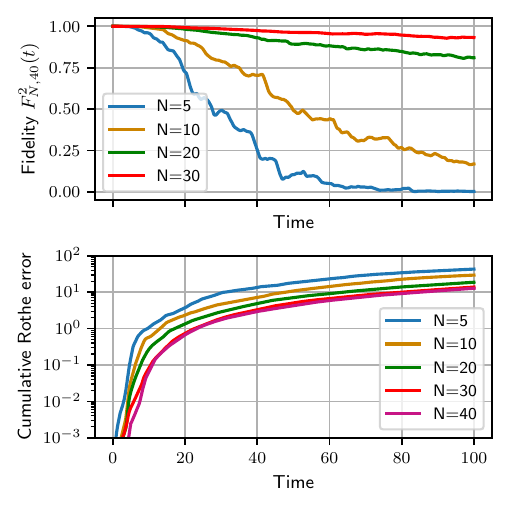}
    \caption{\justifying Error measures for the $2D$ Henon-Heiles potential. Top: Fidelities for $5$, $10$, $20$, and $30$ Gaussians compared to $40$ Gaussians. Bottom: Cumulative Rothe error $r^{\text{cumul.}}(t)$ for $5$, $10$, $20$, $30$, and $40$ Gaussians.}
    \label{fig:2D_fidelities}
\end{figure}

\begin{figure}[h!]
    \centering
    \includegraphics[width=\linewidth]{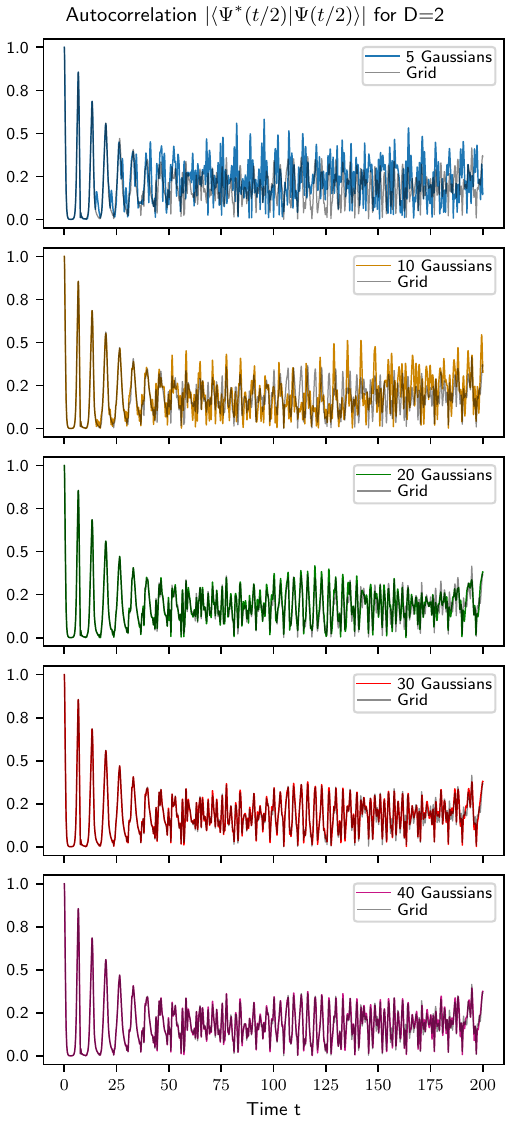}
    \caption{\justifying Autocorrelation functions for the $2D$ Henon-Heiles potential obtained using $5$, $10$, $20$, $30$, and $40$ Gaussians, compared to the grid reference.}
    \label{fig:2D_autocorrelation}
\end{figure}

\begin{figure}[h!]
    \centering
    \includegraphics[width=\linewidth]{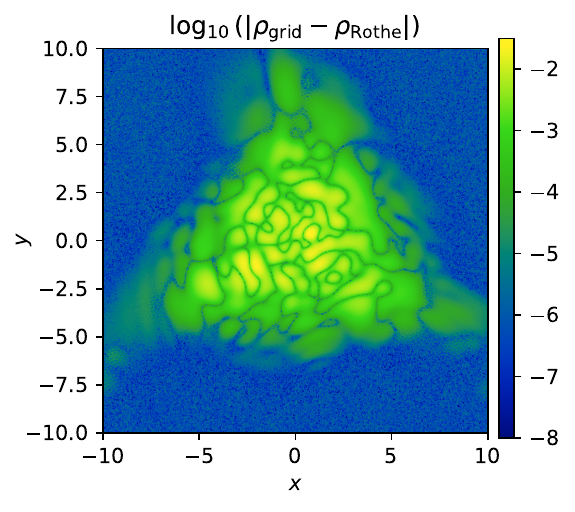}
    \caption{Difference between the density of the grid $\rho_{\rm grid}=|\psi(\boldsymbol r,t=100)_{\rm grid}|^2 $ and $N=40 $ Gaussians propagated with Rothe's method in 2D, $\rho^{\rm Rothe}=|\psi(\boldsymbol r,t=100)_{\rm Rothe}|^2 $.}
    \label{fig:Density_difference}
\end{figure}

Figure \ref{fig:2D_fidelities} shows the fidelities compared to the maximal number of ECGs as a function of the number of ECGs for the $2D$ case. We see that the fidelities for all Gaussians start of at $1$, and that the fidelities decrease more slowly as the number of Gaussians is increased. This shows that the wave functions remain similar to the most accurate solution for longer times with increasing basis set size, for all dimensions considered. In particular, for the $2D$ case, we see that for $N=5$, the fidelity drops to approximately $0.2$ at $t=40$, while for $N=20$, the fidelity remains above $0.8$ at all times. The fidelity for $30$ Gaussians is above $0.92$. While the $N=40$ solution is not completely accurate, this still indicates that as the number of Gaussians is increased, the dynamics converge to the same solution. Comparing this to figure \ref{fig:2D_autocorrelation}, which shows the autocorrelation compared to the grid solution, it also is visible that the dynamics have essentially converged using $N=40$ Gaussians, as the autocorrelation function is very similar to that obtained from the grid calculation.
Looking at the cumulative Rothe errors, however it appears as if the wave functions build up an unreasonably large time evolution error. At time $t=10$, for example, the cumulative Rothe error for $N=5$ Gaussians is above $1$, which indicates that the dynamics can be completely wrong at this time, while it is below $0.1$ for $N=30$ and $N=40$ Gaussians. Nevertheless, the fidelity at that time is $F^2_{5,40}(10)=0.958$, indicating that the two states are still very similar. \newb{In the 2D case, we have also compared the fidelities between the Rothe wave functions and the grid solution at $t=100$ by evaluating the Gaussian based wave functions at the grid points. This is shown in table \ref{tab:fidelities_grid}. In addition, figure \ref{fig:Density_difference} shows the difference between the densities at $t=100$ between the grid solution and the solution using $N=40$ Gaussians. All wavefunctions are normalized before densities and fidelities are computed. As different integrators and different masking procedures were used, we do not expect the two approaches to converge to the same solution - nevertheless, we observe that the fidelity compared to the grid increases with an increasing number of Gaussians to more than $0.9$, and that the difference between the two densities at $t=100$ is not very large, further supporting that the Rothe error overestimates the time evolution error.}
\begin{table}[h!]
    \centering
    \begin{tabular}{ccccc}
        \toprule
        \( N \) & 10 & 20 & 30 & 40 \\ \midrule
        \( F^2_{\text{N,grid}}(t=100) \) & 0.178 & 0.803 & 0.916 & 0.925 \\
        \bottomrule
    \end{tabular}
    \caption{Fidelities \( F^2_{\text{N,grid}}(t=100) \) for \( N=10, 20, 30, 40 \) Gaussians compared to the grid solution.}
    \label{tab:fidelities_grid}
\end{table}

\begin{figure}[h!]
    \centering
    \includegraphics[width=\linewidth]{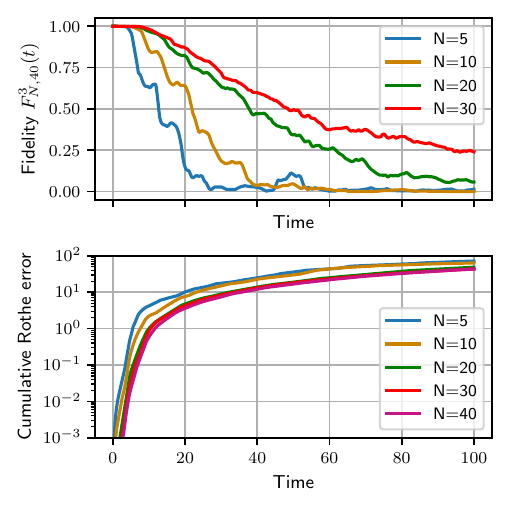}
    \caption{\justifying Error measures for the $3D$ Henon-Heiles potential. Top: Fidelities for $5$, $10$, $20$, and $30$ Gaussians compared to $40$ Gaussians. Bottom: Cumulative Rothe error $r^{\text{cumul.}}(t)$ for $5$, $10$, $20$, $30$, and $40$ Gaussians.}
    \label{fig:3D_fidelities}
\end{figure}
\begin{figure}[h!]
    \centering
    \includegraphics[width=\linewidth]{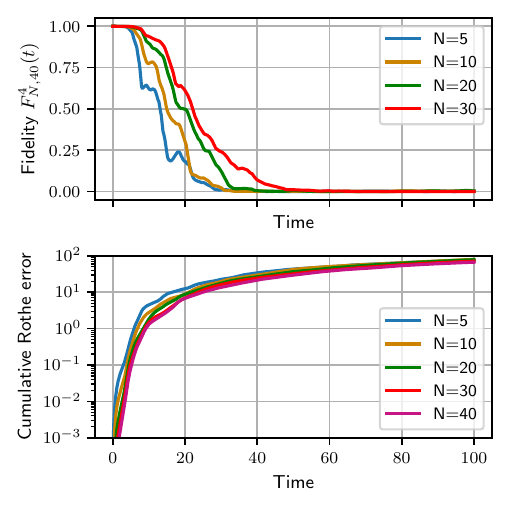}
    \caption{\justifying Error measures for the $4D$ Henon-Heiles potentials. Top: Fidelities for $5$, $10$, $20$, and $30$ Gaussians compared to $40$ Gaussians. Bottom: Cumulative Rothe error $r^{\text{cumul.}}(t)$ for $5$, $10$, $20$, $30$, and $40$ Gaussians.}
    \label{fig:4D_fidelities}
\end{figure}

\begin{figure}[h!]
    \centering
    \includegraphics[width=\linewidth]{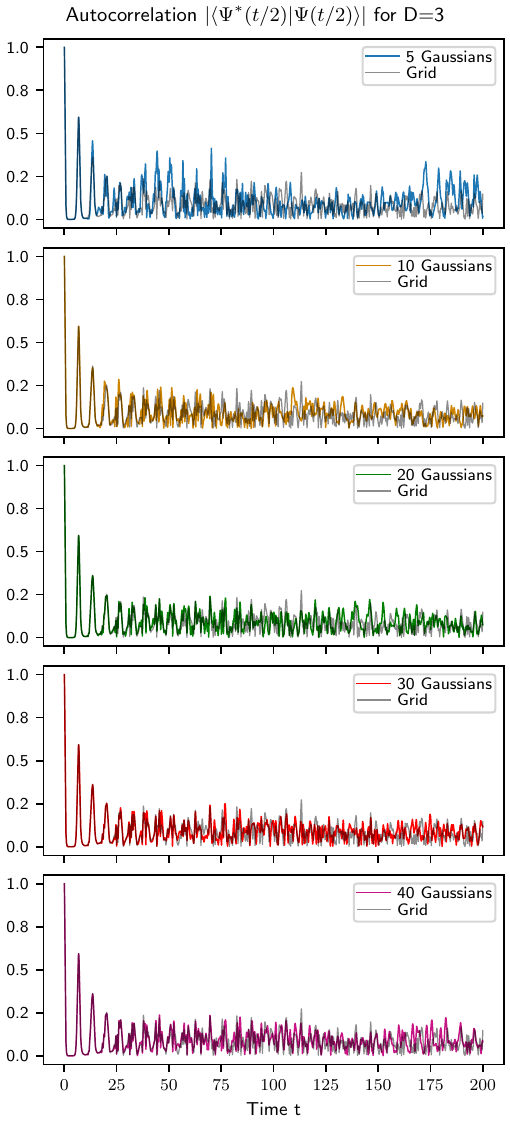}
    \caption{\justifying Autocorrelation functions for the $3D$ Henon-Heiles potential obtained using $5$, $10$, $20$, $30$, and $40$ Gaussians, compared to the grid reference.}
    \label{fig:3D_autocorrelation}
\end{figure}

\begin{figure}[h!]
    \centering
    \includegraphics[width=\linewidth]{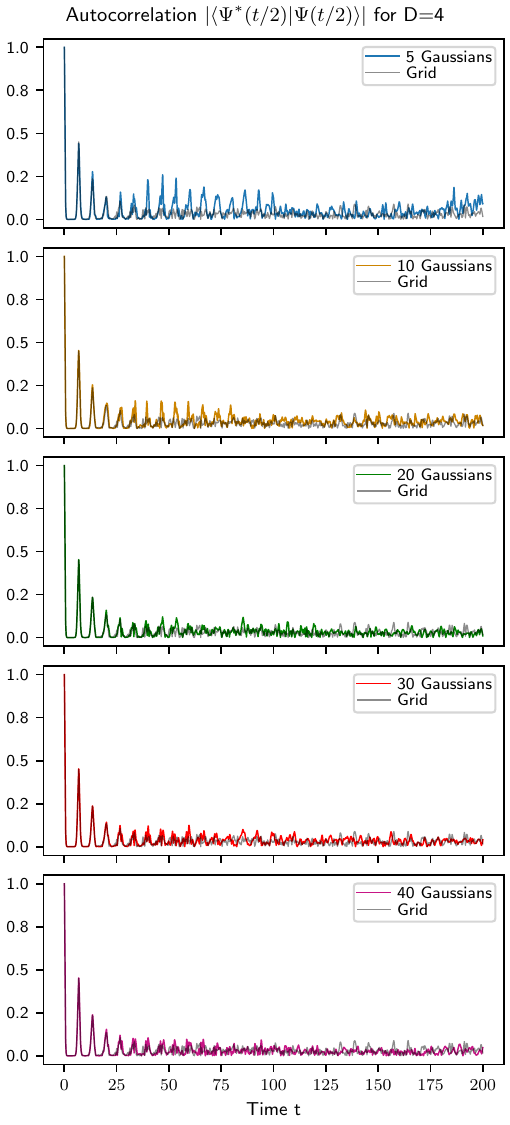}
    \caption{\justifying Autocorrelation functions for the $4D$ Henon-Heiles potential obtained using $5$, $10$, $20$, $30$, and $40$ Gaussians, compared to the grid reference.}
    \label{fig:4D_autocorrelation}
\end{figure}

Similar observations can be made for the 3D and the 4D case, shown in figures \ref{fig:3D_fidelities} and \ref{fig:4D_fidelities}, respectively - the fidelities compared to the calculation with the largest number of Gaussians remain larger for longer times with increasing number of Gaussians, even though the Rothe errors are all very large for all number of Gaussians considered. However, as the fidelities for $N=20$, and $N=30$ in the 3D case, are very small at large $t$, it seems unlikely that the simulations have converged at $t=100$. In the 3D case, we observe that the states using $N=20$ and $N=30$ Gaussians have some overlap with the state using $N=40$ Gaussians, showing that there is some convergence, in the 4D case, however, we observe that from $t=50$ onwards, the fideltities are essentially zero. Similar trends can be observed in the autocorrelation function for the 3D and the 4D Henon-Heiles potentials, shown in figures \ref{fig:3D_autocorrelation} and \ref{fig:4D_autocorrelation}. 

We discuss  and provide an explanation for the discrepancy between the Rothe error and the quality of the fidelity and the autocorrelation in the discussion section.
\subsection{Strang splitting propagation}
\begin{figure}[h!]
    \centering
    \includegraphics[width=\linewidth]{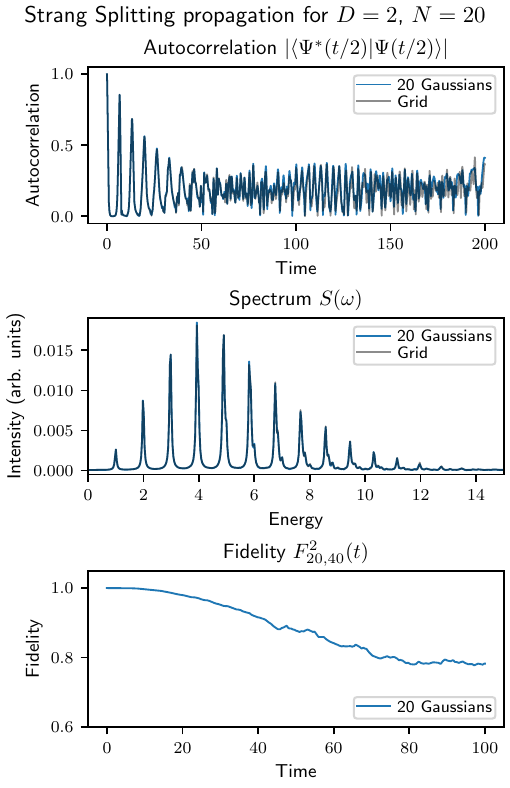}
    \caption{\justifying Autocorrelation, spectrum and fidelity for the 2D Henon-Heiles potential obtained using 20 Gaussians using the alternative propagation scheme. The fidelity is compared to $N_{\text{max.}}=40$, where the original propagation scheme was used.}
    \label{fig:2D_Strang}
\end{figure}

Figure \ref{fig:2D_Strang} shows the autocorrelation function and the corresponding spectrum obtained for $D=2$ dimensions with $N=20$ Gaussians using the Strang splitting approach discussed in section  \ref{sec:StrangSplitting}, in addition to the fidelity compared to the $N=30$ case (where no Strang splitting was used). We observe that the results are qualitatively similar to the $N=20$ case without Strang splitting, with qualitative agreement in both the autocorrelation function, the spectrum, and the fidelity compared to the $N=40$ scheme. 
\section{Discussion}\label{sec:Discussion}
As is expected, we observe an improvement in the quality of the spectra as the number of ECGs is increased. This holds true for all dimensions considered, and we observe for the 2D and 3D plots in particular, that the resulting spectra approach the grid reference. Compared to Ref. \citenum{Burghardt_Worth_HeHe}, who used vMCG with up to 625 frozen, uncorrelated ECGs for the 4D case, we observe that similar results can be achieved with just $\sim\!30$ fully flexible, complex ECGs. In the 2D case, in particular, we have demonstrated that quantitative agreement with the grid solution can be attained, which can be seen from the fidelity with increasing number of Gaussians, as well as the agreement of the autocorrelation function and the spectrum to those obtained on the grid.\\
In the 3D and the 4D case, we obtain spectra that approach the quality of the grid calculation, however, from the fidelities and the autocorrelations, we observe that the wave functions have not converged with $N=40$/$N=30$ Gaussians. Nevertheless, with increased number of Gaussians, the fidelities remain close to one for longer, and the autocorrelations are correct for longer times as well.

For all dimensions considered, we observe that the cumulative Rothe error decreases with increasing number of Gaussians, but it is very large for all simulations presented here, and it remains above $10$ even in the 2D case with $N=40$ Gaussians. Because the Rothe error is large for any number of Gaussians, we did not implement an adaptive scheme to adjust the number of Gaussians. Adding Gaussians whenever the Rothe error is high would result in a prohibitively large number of Gaussians. We have pointed out previously \cite{schrader2024time} that the Rothe error can be much larger than the true time evolution error. In particular, the Henon-Heiles dynamics are chaotic and very hard to reproduce exactly, and the Rothe error indiscriminately measures an upper bound to the deviation from the exact Crank-Nicolson time evolution from $\Psi(t)$ to $\Psi(t+\Delta t)$. One source for the size of the Rothe error is the incorrect representation of the outgoing wave packets, where $\|\hat{H}\Psi\|^2$ is large - even though these dynamics are not of interest.  Minimizing the Rothe error nevertheless works, as it is precisely this minimization that is used in the propagation of the wave function. However, we find fidelities to be a better measure for convergence in the region of interest than the size of the Rothe error: We observe that the times at which the fidelities for different numbers of Gaussians are close to 1, also correspond to the times where the autocorrelation spectra agree with the grid reference. 

We also see that the Strang splitting approach yields comparable results as performing a Rothe propagation with the full Hamiltonian. Due to the fact that many fewer matrix elements need to be calculated, this approach is a feasible alternative in cases when no Gaussians are to be kept frozen, which is generally the case when the initial state is not an eigenfunction of the Hamiltonian - for eigenfunctions, this approach might be counterproductive, as both $\exp(-it\hat{T})$ and $\exp(-it\hat{V})$ introduce non-trivial dynamics, while $\exp(-it\hat{H})$ does not.  
\section{Concluding remarks}\label{sec:Conclusion}
We have demonstrated that Rothe's method can be used to propagate a linear combination of dozens of Explicitly Correlated Gaussians to solve the time-dependent Schrödinger equation in a system with chaotic dynamics. There is no need for any type of regularization that might compromise the quality of the wave function, as is done when using the time-dependent variational principle to propagate Gaussian wavepackets. Furthermore, we have shown that ECGs are extremely flexible functions that are able to well represent the complicated dynamics that arise in the Henon-Heiles potential. Its applicability to quantum dynamics shows that Rothe's method with ECGs can be a viable alternative to the well-established propagation methods such as MCTDH (and its variants) or TDVCC. 
In this work, we explored Rothe's method for the most flexible type of basis functions. However, Rothe's method can be used for any type of parameterization, both with Gaussians or with other basis functions, as long as the matrix elements can be efficiently evaluated. In certain cases, more restricted parameterizations, such as thawed but uncorrelated Gaussians, can offer a viable alternative. In this case, Isserlis' theorem is not required. An indicator for the success of such an approach for many problems is the success of frozen Gaussians in the vMCG method. Similarly, block diagonal correlation matrices would lead to a simplified evaluation of matrix elements.

We have also developed a scheme that allows to use masking functions with ECGs, as well as a simple scheme that leads to norm and energy conservation of the wave function and suggested a scheme that avoids the implementation of most of the squared Hamiltonian $\hat{H}^2$. Further goals include the development of approximation schemes to the Hamiltonian squared and a development of Rothe's method with Gaussians for time-dependent electronic structure theory methods, such as (multiconfigurational) time-dependent Hartree-Fock ((MC)TDHF)\cite{alonUnifiedViewMulticonfigurational2007,MCTDHF1,MCTDHF2}, or Coupled-Cluster (TDCC).\cite{TDCC_review}
Having demonstrated that propagating many fully flexible ECGs with Rothe's method is possible without significant numerical problems, paired with the previous observation that Rothe's method can be used to propagate uncorrelated Gaussians with flexible width to model strong field dynamics of one particle, \cite{schrader2024time} we are hopeful about applying Rothe's method to propagate ECGs for systems with multiple particles without relying on the Born-Oppenheimer approximation \newb{at any stage, \cite{Adamowicz_Kvaal_Lasser_Pedersen} something that is usually done when nuclear motion is considered.\cite{Martinez_review,MCTDH} }The availability of an effective parallelization scheme for Rothe's method suggests that larger systems can be handled.
\section*{Data availability}
The data that supports the finding of this study is available on Zenodo, see Ref.~\citenum{Rothe_zenodo}.
\section*{Author Declaration}
The authors have no conflicts to disclose.

\section*{Acknowledgements}
We thank Professor Ludwik Adamowicz and Mads Greisen Højlund for helpful discussions. The work was supported by the Research Council of Norway through its Centres of Excellence scheme, Project No. 262695. Some of the calculations were performed on resources provided by Sigma2---the National Infrastructure for High Performance Computing and Data Storage in Norway, Grant No. NN4654K.

\bibliography{main}
\end{document}

%% file: intro.tex
In order to understand the dynamics of atomic and molecular systems, it is necessary to solve the time-dependent Schrödinger equation (TDSE). Solving the TDSE is crucial for modeling chemical reactions and understanding the interplay between molecules and light both at the femtosecond scale \cite{FemtoChemistry_paper,FemtoChemistry_book} and at the attosecond scale.\cite{Corkum2007,Krausz2009} Processes and systems that can be modeled and understood by solving the TDSE include, among others, scattering processes, \cite{rescigno2000numerical,newton2013scattering} dissipative systems, \cite{weiss2012quantum} high-harmonic generation (HHG), \cite{HHG_discovery_1,HHG_discovery_2,Lewenstein_model,HHG_methods_review} and vibrational dynamics. \cite{Tannor2007} However, analytical solutions to the TDSE are only available for very simple, idealized systems, making it necessary to use numerical methods. 

A standard approach to solve the time-dependent Schrödinger equation approximately is to choose a parameterization of the wave function and then use one of the time-dependent variational principles (TDVPs),\cite{Dirac1930,Frenkel_wave,McLachlan,kramer1981geometry,Lasser_TDVP} from which equations of motion for the parameters can be derived. A standard high-accuracy method is to represent the time-dependent wave function on a linear basis, i.e. a grid, a discrete variable representation (DVR), or a spline basis. \cite{DVR,boyd2001chebyshev,Bspline_2} While these approaches yield excellent results for a small number of dimensions, their applicability is limited by the exponential scaling in the number of grid points as a function of the dimension. In the context of distinguishable particles, the largest possible dimensionality of these grid-based methods can be extended using the multi-configuration time-dependent Hartree (MCTDH) method, \cite{MCTDH_original,MCTDH,MCTDHbook} where the wave function is represented as a linear combination of Hartree products. Recent grid-based approaches that tackle the dimensionality problems include tensor-network states, \cite{Lubich_TT,Lyu_TT,Greene_TT,Reiher_TT}, and the related multilayer MCTDH (ML-MCTDH) method, \cite{MCTDH_speedup,MLMCTDH,MLMCTDH_TN} in addition to the time-dependent modal vibrational coupled cluster (TDMVCC) method. \cite{Christiansen_VCC_1,Christiansen_VCC_2,Christiansen_TDVCC_1,Christiansen_TDVCC_2,madsen_time-dependent_2020}

A viable alternative to grid-based methods is using a basis of functions where matrix elements can be evaluated analytically. Explicitly Correlated Gaussians (ECGs) have proven to be an invaluable tool for high-resolution calculations of small molecular and atomic systems, both with and without the Born-Oppenheimer approximation, due to their variability, their completeness properties, and in particular the fact that most matrix elements of interest can be calculated analytically. \cite{cafiero2001analytical,Bubin_Adamowicz_ECG2013,Mitroy_Gaussian2013,Adamowicz_Kvaal_Lasser_Pedersen,racsai2024regularized} It has also been shown recently that merely dozens of ECGs can accurately represent two-dimensional wave functions that arise when considering molecules exposed to strong fields. \cite{Wozniak_Gaussians}

When considering dynamics, a historically popular approach is to use a single ECG \cite{Heller_1975} which is propagated using the TDVP either using the full potential or some approximation, such as the Local Harmonic Approximation (LHA).~\cite{Heller_1975} The LHA provides dramatic simplifications, but may give rise to problematic, unphysical behavior. \cite{ryabinkin2024thawed} A single ECG is a relatively crude approximation to the exact wave function, as can be seen from the inherently Gaussian and hence node-less shape of the density. However, this is only the case when a single Gaussian is used, and the quality of the results can be considerably improved by representing the wave function using several ECGs which are propagated independently from one another.\cite{Faou_Lubich_2006,lubich2008quantum,Faou_Lubich_2009, dutra_quantum_2020} A drawback of this approach is that a very large number of ECGs is needed to properly represent the wave function, as independent propagation does not lead to a compact representation.

In the variational Multi-Configurational Gaussian method (vMCG) and the related Gaussian MCTDH method (G-MCTDH),\cite{G-MCTDH,vMCG} the wave function is (partially) represented as a linear combination of Gaussian wave packets, and the linear and nonlinear Gaussian parameters are updated inter-dependently using the TDVP. However, numerical instabilities may arise due to the ill-conditioning of the Gramian matrix, which has to be inverted at every time step when solving the equations of motion.~\cite{Sawada,G-MCTDH,vMCG,KAY1989165,Lee2018,HaakonPhD} There are many successful approaches to overcome this issue, such as using uncorrelated or frozen Gaussians with a fixed width,~\cite{Heller_1981,Vanicek2020,Burghardt_Worth_HeHe,vMCG} specific initialization schemes, independent propagation schemes, \cite{dutra_quantum_2020} regularization schemes, \cite{vMCG,G-MCTDH} or the LHA. Combined with a proper choice of the initial Gaussians, combinations of these schemes can give very good results for large systems.\cite{Burghardt_ref1,Burghardt_ref2,Burghardt_ref3,Burghardt_ref4,Burghardt_ref5}
However, the basis set size required increases when Gaussians are frozen or uncorrelated, and regularization and independent propagation can lead to uncontrolled errors. \cite{feischl2024regularized}

Recently, Rothe's method\cite{Rothe1930_MA,deuflhard2012adaptive} has been used to propagate linear combinations of Gaussians with variable width and shift parameters to study the dynamics of a one- and a three-dimensional Hydrogen atom exposed to a strong laser pulse. \cite{kvaal2023need,schrader2024time} In Rothe's method, time evolution is rephrased as an optimization problem, and many of the numerical issues that arise when propagating ECGs with a variational method are circumvented. To test the applicability of Rothe's method to propagate linear combinations of ECGs in a challenging situation, we consider the Henon-Heiles potential. The Henon-Heiles potential is a test potential where there are both long-lived resonant states as well as quickly outgoing unbound states, with harmonic behavior close to the origin and anharmonic behavior further away, and very complicated ensuing chaotic dynamics.\cite{NordholmRice_HeHe,NoidMarcus_HeHe,DavisHeller_HeHe,Bixon_HeHe,Feit_HeHe,Zuniga_HeHe} The Henon-Heiles potential has been used as a test potential to verify the applicability of numerical methods, including semiclassical methods, \cite{DavisHeller_HeHe,Walton_Manolopoulos_HeHE,Wang_HeHe,BrewerHeHe,Child_HeHe} MCTDH,\cite{MCTDH_original,NestHeHe} vMCG, \cite{Burghardt_Worth_HeHe}, ML-MCTDH, \cite{mlmctdh_hehe} and TDVCC. \cite{Christiansen_TDVCC_1}

In this paper, we consider the dynamics of an initially Gaussian wave packet in the Henon-Heiles potential, where the wave function is written as a linear combination of fully flexible multidimensional ECGs and propagated using Rothe's method. In section \ref{sec:Methods}, we present the theory regarding explicitly correlated Gaussians and describe both variational dynamics for single ECGs as well as Rothe's method to propagate an arbitrary number of ECGs, focusing particularly on the calculation of matrix elements, the use of mask functions and optimization of the nonlinear parameters. In section \ref{sec:system-implementation}, we describe the model Hamiltonian and the initial conditions, the results we present, the reference schemes which we compare our results to, and some computational details. In section \ref{sec:Results}, we present the results obtained using Rothe's method and compare them to our reference calculations, followed by discussion in section \ref{sec:Discussion}. We conclude with a summary and future outlooks in section \ref{sec:Conclusion}.